\documentstyle[12pt]{article}

\documentstyle{titlepage}
\title{DIRECT PARTICLE INTERACTION AS THE ORIGIN OF THE QUANTAL BEHAVIOURS}
\author{\bf ALI\ SHOJAI$^*$\ \&\ MEHDI\ GOLSHANI$^{**}$\\
Department of Physics, Sharif University of Technology\\P.O.Box 11365-9161 Tehran, IRAN\\
and\\Institute for Studies in Theoretical Physics and Mathematics,\\P.O.Box 19395-5531, Tehran, IRAN\\
$^*$Email: SHOJAI@PHYSICS.IPM.AC.IR\\
$^{**}$Fax: 98-21-8036317\\}
\date{}
\begin{document}
\begin{bf}
\maketitle
\vspace{1cm}
\begin{center}
{\Large DIRECT PARTICLE INTERACTION AS THE ORIGIN OF THE QUANTAL BEHAVIOURS}\\
{\bf A. Shojai \& M. Golshani}
\end{center}
\vspace{0.5cm}
\begin{center}
{\bf ABSTRACT}
\end{center}
{\it It is argued that the quantal behaviours may be understood
in the framework of direct particle interactions. A specific example
is introduced. The assumed potential predicts
that at sufficiently large distances quantal behaviours arise, while at very
large distances gravitational-like forces are present. The latter is true provided
all particles have internal structures.}
\vspace{1.5cm}
\section{Direct Particle Interaction Versus Field Theory}
\hspace{0.5cm}In formulating his ideas and those of former physicists about
motion and gravitation, Newton made two fundamental assumptions:\\
(a)-- He postulated the presence of {\it absolute space and time} --- physical objects 
which act on the particles, but are not acted upon by them.\\
(b)-- He assumed that absolute space is isotropic and homogeneous and that absolute
time is also homogeneous. In other words, he supposed that physics is invariant under 
Galileo's transformations.\\
An important result of these assumptions is that the particles interact at a distance, i.e.
information propagates with infinite velocity[1]. Theories containing 
{\it action-at-a-distance\/} (AAAD) interactions are examples of {\it direct particle interaction\/} (DPI)
theories.
\par
Some people felt uncomfortable with the above assumptions. Philosophers like
Berkeley, Leibnitz, Mach and others, argued that the only physically meaningful
thing for moving particles is the {\it relative motion\/} and thus it is difficult
and unnecessary to believe in absolute space and time. It is possible to show
that such a relational physics must be invariant under Leibnitz transformations[2]:
\begin{equation}
\left \{ \begin{array}{l} \vec{x} \rightarrow \stackrel{\leftrightarrow}{A}(t)\cdot \vec{x} + \vec{B}(t) \\
t \rightarrow C(t) \end{array} \right .
\end{equation}
where $\stackrel{\leftrightarrow}{A}(t)$ is an orthogonal matrix and $\vec{B}(t)$ and $C(t)$ are
arbitrary functions of time. Barbour et.al.[2-5] have constructed lagrangians which
are Leibnitz-invariant and shown that locally one arrives at our standard physics.
They are also able to relate local parameters to cosmological ones in their theory.
\par
The apparent contradiction between Maxwell's electromagnetic theory and Newton's
theory of motion, led some people to doubt the absoluteness of space and time
and to the postulation of absolute space-time. In this way the special theory of relativity 
and Lorentz transformations were born. Later when Einstein tried to bring
gravitation in this framework, he was forced to throw out the absoluteness of space-time but not its existence.
An important result of relativity theory is that the velocity of particles 
as well as the information propagation velocity cannot exceed a universal value ---
the velocity of light. Accordingly, there are two possible ways of describing the interaction 
between particles. First, one can introduce the {\it field\/} concept, a physical
object which propagates with a finite velocity which is less than or equal to that of light.
All information between particles is carried by the field. Second, it is possible
to look at the situation through the glasses of DPI. Particles interact
with each other directly but not instantaneously. In the case of electromagnetism, where
the velocity of propagation of information is equal to that of light, we have
{\it action-at-a-zero-proper-distance\/} (AAAZPD). The first suggestion is what actualy
is used in Maxwell's theory of electromagnetism. Schwarzschild, Tetrode, Fokker,
Wheeler, Feynman, Hoyle, and Narlikar[6-18] developed a DPI theory for 
electromagnetism which produces all of the results of Maxwell's theory and in addition
predicts two important things --- the self force and the existance of only the retarded
solutions.
\par
On the other hand, quantum mechanics brought with itself idealism,
indeterminism and nonlocality. The first two have
been overcome in the beautiful theory of Bohm[19-25]. He showed how one can remain faithful
to realism and determinism as in classical physics and at the same time be able
to reproduce all of the results of quantum mechanics, via introducing a {\it quantum potential\/} (QP) in
Newton's equation of motion. In this way, quantum phenomena are nothing but
physical situations in which a new force, the quantum force, derivable from the QP:
\begin{equation}
Q=-\frac{\hbar ^2}{2m} \frac{\nabla ^2 \sqrt{\rho}}{\sqrt{\rho}}
\end{equation}
is present. In the above relation $\rho (\vec{x},t)$ is the density of an ensemble
of particles. The QP has the peculiar property that it is not dependent on 
the magnitude of density, it is a function of the shape of the square root of it.
The nonlocality of quantum mechanics (i.e. the presence of AAAD) which can be seen 
both through the QP[25,26] and in Bell's theorem[27], has apparently been proved experimentally[28].
Therefore one is forced to set some limitations on the validity domain of the 
relativity theory. The QP always acts via AAAD. It cannot be formulated either as an
AAAZPD or as a field theory. Although some works have been done, to
make Bohm's theory consistent with relativity[29-31], none of them are
acceptable, either because of theoretical problems or because of the
lack of agreement with experiments.
\par
In addition to this apparent contradiction between relativity and quantum theories,
there is yet another problem. When one combines these theories and applies it to  
Maxwell's theory, one gets some amazing results. In Maxwell's theory the interaction 
between charged particles is transported by the electromagnetic field, while 
in quantum electrodynamics, this interaction is mediated by 
{\it particle-like-states\/} called {\it photons\/}. It is always stated that 
photons travel at light's speed and thus there is no room for mysterious AAAD.
But, we must note that first of all, the virtual photons are not on the mass shell,
so they may have any velocity from zero to infinity. Although when one sums over 
all possible paths, the result is Lorentz-invariant, this seems to be in contradiction with the spirit
of relativity theory. Second, although quantum field theory removes the need for AAAD,
it leads to infinities. Investigation of some of these infinities (by the point splitting renormalization method, say)
shows that they are the results of interaction of photons and charged particles at 
a point ({\it action-at-a-point\/} or AAAP). To avoid them, one must let the
interaction takes place at a distance! In summary one can choose either 
AAAD or AAAP, but the latter leads to infinities.
\par
Summing up our discussion, one is forced to accept that the
correct physical theory must be relational and containing DPI. We 
ruled out Galileo and Lorentz transformations and chose Leibnitz transformations
because of two facts. First, we know that they are only of limited validity,
and, as Barbour et.al.[2-5] have shown, they are local approximations
of Leibnitz transformations. Second, a relational physics rules out the unphysically 
existing self-dependent space-time.
\par
In the following, a specific and appropriate DPI is suggested, using the general properties 
of DPIs. This typical DPI is founded on a trivial, tautological
postulate. It contains two scale factors, a short scale $\alpha _s$ and a large scale
$\alpha _{\ell}$. It will be shown that at distances larger than $\alpha _s$,
this prototype DPI is equal to the Bohm's QP plus some small corrections.
Therefore it provides a framework for understanding the mysterious quantal
behaviours in terms of instantaneous interaction between different particles of the ensemble.
As it is wellknown[25], the QP plus the nutural constriant on density to obey the continuity equation, leads to the schrodinger equation.
So in fact we shall derive the quantum theory from DPI.
An important property of our prototype DPI is that its small corrections to the QP 
magnify the internal structure of any particle at large distances. They lead to gravitational-like forces.
Therefore it is suggested that DPI theories are suitable for unifying gravity and quantum mechanics (which
from Bohm's point of view is nothing but a fifth force).
A good DPI theory must unifiy at the same time all of the five forces (gravity, electromagnetism, weak, strong and quantum forces).\\

\section{A Typical DPI}
\hspace{0.5cm}As it was discussed in the previous section, many areas of physics, 
including quantum mechanics and Newtonian gravity, are understandable in terms of AAAD, i.e. nonlocality.
It is also argued that DPI theories seem to be a natural framework for describing nonlocal phenomena 
and thus for unifying different parts of physics. In this section we shall develop
a typical DPI, and later, in the forthcoming sections, we show that under certain conditions
it reduces to the QP or to the Newtonian gravity.
\par
To begin with, let us stress a trivial property of DPIs. It is clear that any DPI,
highly depends upon the configuration of particles, i.e. on their relative position.
Therefore the first task in constructing any DPI is to ensure that each particle is at its correct position, i.e.
the position derived from the equation of motion. Accordingly we postulate 
the following tautological statement:\\
\newline
{\bf Postulate}: {\it Each particle is at its own location\/}.
\newline
\par
It seems unbelievable that this postulate can lead to any physical consequences,
but as we shall see, it is essential in obtaining the QP.
\par
Now let us formulate this postulate. Consider a system of $N$ identical
particles, each one located at $\vec{a}_i(t)$, $i=1\cdots N$.
One can imagine that this 
pattern of particles is made by bringing particles in one by one, and locating them
at their correct position. In order to ensure that each particle is 
at its right position, the DPI potential should contain a factor, which is infinitely
large when some particle is at incorrect position and is finite otherwise.
This can be achieved for each particle, if we make use of the Dirac delta function, in the form $1/\delta (\vec{x}-\vec{a}_i(t))$.
Since we assume that all particles are identical, each particle may be put at 
any of $\vec{a}_i(t)$'s. So the corresponding factor in the DPI
potential is $1/\sum _{i=1}^N \delta (\vec{x}-\vec{a}_i(t))$.
This is zero when $\vec{x}$ is equal to some legal position and is infinite
elsewhere. 
\par
Apart from this factor, the DPI potential may contain a factor
(obviously relational) representing relative configuration of particles.
To have a definite model, we assume two kinds of interactions: a short range
interaction and a long range one with ranges $\alpha _s$ and $\alpha _{\ell}$ respectively.
In addition, we assume these interactions be exponential. Therefore, as a typical
DPI potential, we consider the following one:
\begin{equation}
U(\vec{x},t)=\frac{U_0}{\sum _{i=1}^N \delta (\vec{x}-\vec{a}_i(t))} 
\left \{ \sum_{j=1}^N \exp [-(\vec{x}-\vec{a}_j(t))^2/\alpha _s^2] \right \}
\left \{ \sum_{k=1}^N \exp [-(\vec{x}-\vec{a}_k(t))^2/\alpha _{\ell}^2] \right \}
\end{equation}
where $U_0$ is some constant. We shall work with this DPI potential throughout this paper.
Two notes must be remarked here. First, the first
factor is equal to $1/\rho (\vec{x},t)$. This says the following: {\it particles like to go where
particles are present}. Second, the exponential form is not necessary. In fact,
if these terms fall faster than $1/x^2$ all of the forthcoming results can be obtained.
We choose this form for simplicity. In other parts of this work we shall show that
the QP and the Newtonian gravitational potential are derivable from this prototype DPI potential.\\

\section{QP As A Result Of DPI}
\hspace{0.5cm}In this section, we use our prototype DPI to derive the Bohm's QP. Suppose we are
dealing with particle separations larger than $\alpha _s$. Multiply $U(\vec{x},t)$ by the identity factor 
$1=(\pi \alpha _s^2)^{3/2}/(\pi \alpha _s^2)^{3/2}$, and use the identity:
\begin{equation}
\int d^3y \exp \{ -[\vec{y}+\beta (\vec{x}-\vec{a}_k(t))]^2/\alpha _s^2 + y^2/\gamma ^2 -y^2/\gamma ^2\} = (\pi \alpha _s ^2)^{3/2}
\end{equation}
for arbitrary $\beta$ and $\gamma$. By choosing $\beta$ and $\gamma$ in the form:
\begin{equation}
\beta=\frac{1}{2}\left [ 1+\sqrt{1-4\alpha _s^2/\alpha _{\ell}^2} \right ]
\end{equation}
\begin{equation}
\gamma=\frac{\sqrt{2}\alpha _s }{\left [ 1-\sqrt{1-4\alpha _s^2/\alpha _{\ell}^2} \right ] ^{1/2}}
\end{equation}
and after a little algebra and noting that for small $\epsilon$ (or in other words for $(\vec{x}-\vec{a}_k(t))^2 \gg \epsilon^2$) we have
the following representation for the square root of Dirac's delta function:{\footnote{Note
that we have used the parameter $\alpha_s$ as the small parameter of the representation 
of Dirac's delta function. It may seem that it is an arbitrary choice, but it can be seen
that if one chooses another small parameter, the result is of the same form as in (21) with different 
coefficients. The above choice is the most economical one.}}

\begin{equation}
\left ( \frac{1}{\pi \epsilon^2} \right )^{3/2} \exp [-(\vec{x}-\vec{a}_k(t))^2/\epsilon^2] \simeq \delta(\vec{x}-\vec{a}_k(t))
\end{equation}
with $2\epsilon^2=\alpha_s^2$ and using the following fact:
\begin{equation}
\sum_{k=1}^N \sqrt{\delta(\vec{x}-\vec{a}_k(t))} = \sqrt{\sum_{k=1}^N \delta(\vec{x}-\vec{a}_k(t))}
\end{equation}
which can be proved by using the step function representation of Dirac's delta function, and using the definition of the density of particles:
\begin{equation}
\rho(\vec{x},t)=\sum_{k=1}^N \delta(\vec{x}-\vec{a}_k(t))
\end{equation}
one can easily show that the DPI potential can be written as:
\begin{equation}
U(\vec{x},t)\simeq U_0 (4\beta)^{-3/4} 
\int d^3y \sqrt{\frac{\rho(\vec{x}+\vec{y},t)}{\rho(\vec{x},t)}} \exp \left [ -\frac{y^2}{\gamma^2} \right ]
\end{equation}
This is an equivalent form of equation (3) for our typical DPI, written in terms of the density
of the ensemble. But it must be noted that equations (10) and (3) are the same, only
for $(\vec{x}-\vec{a}_k)^2\gg \alpha_s^2$. Thus we assume that the correct potential, both for
small and large separations is (3).
\par
Our aim is now, to show the relation between (3) or (10) and
Bohm's QP. In order to do this, we express $\sqrt{\rho(\vec{x}+\vec{y},t)}$ in terms
of (7)--(9). The integral is then Gaussian and can be carried out:
\begin{equation}
U(\vec{x},t)\simeq U_0 \left ( \frac{1}{4\pi \epsilon ^2 \beta}\right )^{3/4} \left ( \frac{\pi}{1/2\epsilon^2+1/\gamma^2}\right )^{3/2}
\frac{\sum_{k=1}^N \exp \left [ -\frac{(\vec{x}-\vec{a}_k(t))^2}{2\epsilon^2+\gamma^2}\right ] }{\sqrt{\rho(\vec{x},t)}}
\end{equation}
Note that for $\alpha_l\gg \alpha_s$ this is proportional to:
\[ \frac{\sum_{k=1}^N \exp \left [ -(\vec{x}-\vec{a}_k(t))^2/\alpha_l^2 \right ]}{\sum_{k=1}^N \exp \left [ -(\vec{x}-\vec{a}_k(t))^2/2\epsilon^2\right ]} \]
This is a form which can be obtained directly from equation (3) by using the above Gaussian representation
of Dirac's delta function.
\par
Equation (11) is a form for DPI that contains both the density and the particles' position. Relation to Bohm's QP
would be appear if the DPI is written in terms of the density, only. In order 
to arrive at this aim, we use Backer--Hausdrof lemma:
\begin{equation}
e^{-G}Ae^G=A-[G,A]+\frac{1}{2}[G,[G,A]]+\cdots
\end{equation}
and set $G=G(\vec{x})$ and $A=\vec{\nabla}$. Nothing that
\begin{equation}
[G,\vec{\nabla}]=-\vec{\nabla}G,\ \ \ \ \ \ [G,[G,\vec{\nabla}]]=0\ \ \ \ etc.
\end{equation}
we have
\begin{equation}
e^{-G}\vec{\nabla}e^G=\vec{\nabla}+\vec{\nabla}G
\end{equation}
This is an operator identity. Now suppose that it acts on the unity:
\begin{equation}
\left ( e^{-G}\vec{\nabla}e^G\right )1=\left (\vec{\nabla}G\right )1
\end{equation}
If one uses this relation for the following operator:
\begin{equation}
e^{\omega \nabla ^2} \equiv 1+\omega \nabla ^2 +\cdots
\end{equation}
one has:
\begin{equation}
\left ( e^{-G}e^{\omega \nabla ^2}e^G\right )1=e^{\omega |\vec{\nabla}G|^2} 
\end{equation}
In this relation we choose:
\begin{equation}
G=-\frac{(\vec{x}-\vec{a}_k(t))^2}{2\epsilon^2}
\end{equation}
So:
\begin{equation}
e^{\omega \nabla ^2}e^{-(\vec{x}-\vec{a}_k(t))^2/2\epsilon^2}=e^{(-1/2\epsilon^2+\omega /\epsilon^4)(\vec{x}-\vec{a}_k(t))^2}
\end{equation}
If one sets:
\begin{equation}
\omega=\frac{1}{2}\frac{\epsilon^2\gamma^2}{2\epsilon^2+\gamma^2}
\end{equation}
the relation (11) can be simplified as:
\[ U(\vec{x},t)\simeq U_0 \left ( \frac{1}{4\pi \epsilon ^2 \beta}\right )^{3/4} \left ( \frac{\pi}{1/2\epsilon^2+1/\gamma^2}\right )^{3/2}
\frac{e^{\omega \nabla ^2}\sum_{k=1}^N \exp \left [
-\frac{(\vec{x}-\vec{a}_k(t))^2}{2\epsilon^2}\right ] }{\sqrt{\rho(\vec{x},t)}} \]
\[ = U_0 \left ( \frac{1}{4\pi \epsilon ^2 \beta}\right )^{3/4} \left ( \frac{\pi}{1/2\epsilon^2+1/\gamma^2}\right )^{3/2}
\frac{e^{\omega \nabla ^2}\sqrt{\rho}}{\sqrt{\rho}} \]
\begin{equation}
=U_0 \left ( \frac{1}{4\pi \epsilon ^2 \beta}\right )^{3/4} \left ( \frac{\pi}{1/2\epsilon^2+1/\gamma^2}\right )^{3/2}
\frac{1}{\sqrt{\rho}} \left ( 1+\omega \nabla^2 \sqrt{\rho}+\cdots \right )
\end{equation}
But this is just Bohm's QP corrected by small terms! 
Thus, conclusion is that at separations larger than $\alpha _s$, our DPI leads to Bohm's QP with some corrections. Note that the approximate nature of (21)
is due to (7).
\par
At this point it is worthwhile to note that our DPI potential in form (3)
seems to be large when $\rho$ is small, but this property is not apparant in (21).
A glance at the derivation of (21) from (3) shows that the exponential terms
in (3) are related to $\rho$, and thus in taking the limit of small $\rho$, their  
role must be considered. In fact the case of small densities needs some caution. (See e.g. [25])\\
\section{Observations}
\hspace{0.5cm}We have seen that Bohm's QP, may be viewed as a result of a
DPI. Here is some points:\\
\newline
(a)-- In the previous section, we see that for a system of $N$ similar particles,
our prototype DPI led to the QP. In our derivations an essential assumption was made. It was
assumed that the density function $\rho (\vec{x},t)$ is a differentiable
function of $\vec{x}$. This is true only for large $N$. Therefore the prototype
DPI leads to the QP for a large ensemble of particles.\\
\newline
(b)-- Although the QP may be derived for an ensemble of similar particles, it is
also the correct potential for a particle observed a large number of times.
The term representing our tautological postulate must be interpreted as follows. The DPI potential
has to be finite when the particle is at any allowed position and infinite elsewhere.
The exponential terms have similar interpretations.\\
\newline
(c)-- The above one-particle derivation may be generalized to the many-particle case.
Consider $N$ particles of kind $1$, $N$ particles of kind $2$, $\cdots$, and $N$ particles
of kind $m$. Then the DPI potential must be written as:
\[ U(\vec{x}_1,\cdots ,\vec{x}_m,t)=\frac{U_0}{\sum _{i=1}^N \prod _{l=1}^m\delta (\vec{x}_l-\vec{a}_i^{(l)}(t))} \]
\begin{equation}
\times \left \{ \sum_{j=1}^N \prod _{l=1}^m\exp [-(\vec{x}_l-\vec{a}_j^{(l)}(t))^2/\alpha _s] \right \}
\left \{ \sum_{k=1}^N \prod _{l=1}^m\exp [-(\vec{x}_l-\vec{a}_k^{(l)}(t))^2/\alpha _{\ell}] \right \}
\end{equation}
where $\vec{a}_i^{(l)}(t)$ represents a legal position for a particle of kind $l$.
In a manner very similar to that of the one-particle case, one can show that the DPI
potential can be written as:
\[ U(\vec{x}_1, \cdots ,\vec{x}_m,t)\simeq U_0 (4\beta)^{-3m/4}\left (\frac{\pi}{1/2\epsilon^2+1/\gamma^2}\right )^{3m/4} \frac{1}{\sqrt{\rho (\vec{x}_1, \cdots ,\vec{x}_m,t)}} \]
\begin{equation}
\times \exp \left \{\omega [\nabla _1^2+\cdots +\nabla _m^2] \right \} \sqrt{\rho (\vec{x}_1, \cdots ,\vec{x}_m,t)}
\end{equation}
for separations larger than $\alpha _s$ and for large $N$. This is just Bohm's
expression for the QP of a many-particle system, corrected by small terms.
Now, although we have derived the above expression for an ensemble of similar many-particle systems, it is also true for 
any many-particle system.\\
\newline
(d)-- As it was seen in the previous section, our prototype DPI potential is equivalent
to Bohm's QP with some small corrections. Now we show that this correction terms
have the property of magnifying the small scale structure of matter to large 
distances. Since in the expression (21), derivatives of very high degree exist, very far
points are connected. To see this, suppose we discretize the space by unit $\epsilon$,
and consider $\rho (\vec{x},t)$ to be spherically symmetric. Then we have:
\begin{equation}
\nabla ^2 \sqrt{\rho}\simeq \frac{\sqrt{\rho (r-2\epsilon)}}{\epsilon ^2} + \ other \ \ terms.
\end{equation}
So the DPI potential may be written as:
\[ U(\vec{x},t)\simeq U_0 (4\beta)^{-3/4}\left (\frac{\pi}{1/2\epsilon^2+1/\gamma^2}\right )^{3/4} \frac{1}{\sqrt{\rho(r)}} \]
\begin{equation}
\times \sum _{n=0}^{M/2-1} \frac{1}{n!} \omega ^n \frac{1}{\epsilon ^{2n}} \sqrt{\rho(r-2n\epsilon)} + \ other \ \ terms.
\end{equation}
where
\begin{equation}
M=\frac{r}{\epsilon}
\end{equation}
If we choose
\begin{equation}
\epsilon = \sqrt{\omega}\simeq \frac{\alpha _s}{2}
\end{equation}
we obtain:
\begin{equation}
U(\vec{x},t)\simeq U_0 (4\beta)^{-3/4}\left (\frac{\pi}{1/2\epsilon^2+1/\gamma^2}\right )^{3/4} \frac{1}{\sqrt{\rho(r)}} 
\sum _{n=0}^{M/2-1} \frac{1}{n!} \sqrt{\rho((1-2n/N)r)} + \ other \ \ terms.
\end{equation}
Note that this expression is acceptable only for 
\begin{equation}
r \gg \epsilon \sim \alpha _s
\end{equation}
As it is seen in this relation, the DPI potential
relates any point, to the density at a distant location.
\par
As a model, suppose that the universe is made of a uniform distribution of matter with density $\rho _0$
and a particle with very fine internal structure like below:
\begin{equation}
\rho(r)=\zeta ^2 \left [1-e^{-\xi/r}\right ]^2
\end{equation}
This is a sharp function of $r$ provided $\xi$ is very small.
The DPI potential is now:
\begin{equation}
U(\vec{x},t)\simeq \frac{U_0 (4\beta)^{-3/4}\left (\frac{\pi}{1/2\epsilon^2+1/\gamma^2}\right )^{3/4}}{\sqrt{\rho(r)}} 
\zeta \xi \left (\sum _{n=0}^{M/2-1} \frac{1}{n!(1-2n/N)}\right ) \frac {1}{r} + \ other \ \ terms.
\end{equation}
This is just the Newton's law of gravitation plus some corrections.
Note that it has the correct sign and that there is a relation between 
the Plank's constant, Newton's constant of gravitation, the density of universe and the parameters of the internal 
structure of particles.
\par
In summary, if we assume that all particles have internal structures below the $\alpha _s$ scale,
at large distances ($r \gg \alpha _s$) one sees some gravitation-like forces!\\
\newline
(e)-- To complete our discussion, we must consider the kinetic terms and study the dynamics
of the system. Barbour et. al.[2-5] have shown that Leibnitz-invariant lagrangians
are of the form of the product of a kinetic term ${\cal K}$ and a potential term ${\cal P}$:
\begin{equation}
{\cal L}={\cal K}{\cal P}
\end{equation}
In our case, for a one-particle system, the potential term is
\begin{equation}
{\cal P}=\sum _{i=1}^N U(\vec{a}_i(t))
\end{equation}
while we have chosen the kinetic term to be
\begin{equation}
{\cal K}=\left ( \sum _{i=1}^N \left [ \frac{d}{dt} \mid \vec{a}_i(t)-\vec{a}_j(t) \mid \right ]^2 \right )^{1/2}
\end{equation}
Barbour and others[2] have shown that the exponent one-half is necessary for the
action be Leibnitz-invariant. In accordance with their work, one is able to relate the local physics to 
cosmology. In this way the coupling constant of the quantum force, i.e.
$- \hbar ^2/2m$ is related to the cosmological parameters like the radius of the
universe, its the expansion velocity, its density and so on. Therefore $\hbar$
may be a function of time depending on our choice of cosmological model.\\
(f)-- As a cosmological model, Consider the universe made of a shell of radius
$R$, moving radialy with velocity $\dot{R}$, as well as $N$ particles located at $\vec{a}_i(t)$. 
Then the kinetic term can be written as sum over shell-shell points plus
sum over shell-particles plus sum over particle-particle. The result is (see [2] for a similar calculation):
\begin{equation}
{\cal K}= \ constant\ \left [ 1+\frac{1}{30 \dot{R}^2}\sum _{i=1}^N \left (\frac{d\vec{a}_i}{dt}\right )^2 + \cdots \right ]
\end{equation}
So the lagrangian is approximately:
\begin{equation}
{\cal L} \simeq \ constant \ \left \{ \frac{1}{2}m \sum _{i=1}^N \left (\frac{d\vec{a}_i}{dt}\right )^2 -\frac{\hbar ^2}{2m} \sum _{i=1}^N\left (\frac{\nabla ^2 \sqrt{\rho}}{\sqrt{\rho}}\right )_{\vec{x}=\vec{a}_i} 
-Gm \sum _{i=1}^N \frac{1}{\mid \vec{a}_i-\vec{a}_j \mid} + \cdots \right \}
\end{equation}
where we have:
\begin{equation}
\frac{\hbar (t)}{\hbar(t_0)}=\frac{\dot{R}(t)}{\dot{R}(t_0)}
\end{equation}
and:
\begin{equation}
\frac{G(t)}{G(t_0)}=\sqrt{\frac{\rho _{universe}(t_0)}{\rho _{universe}(t)}}
\end{equation}
\section{Conclusion}
\hspace{0.5cm}It is shown that the DPI is a natural framework for quantal phenomena.
As our prototype DPI potential shows, one is able to describe both the QP and the gravity
as different aspects of a single interaction, provided that all particles have internal structure.
Thus we hope that different parts of physics may be different aspects of a specific
DPI, although the construction of such a DPI model needs a large amount of work.\\
\newline
{\bf Acknowledgement:} We are grateful to Prof. B. Mashhoon for fruitful discussions.
We also thank F. Shojaei for her fruitful remark.\\
\newline
\newline
{\bf References\\}
[1]-- Landau, L.D., and E.M. Lifshitz, {\it The classical theory of fields\/},
Addison-Wesley, Reading, Mass., and Pergamon press, Oxford, England, (1962).
\newline
[2]-- Barbour, B.J., and B. Bertotti, Il Nouvo Cimento, {\bf 38B}, No. 1, 1, (1977).
\newline
[3]-- Barbour, B.J., Nature, {\bf 249}, 328, (1974).
\newline
[4]-- Barbour, B.J., Il Nouvo Cimento, {\bf 26B}, No. 1, 16, (1975). 
\newline
[5]-- Barbour, B.J., and B. Bertotti, Proc.  Roy. Soc. Lond., {\bf 382A}, 295, (1982).
\newline
[6]-- Schwarzschild, Gottinger Nachrichten, {\bf 128}, 132, (1903).
\newline
[7]-- Tetrode, H., Zeits. f. Physik, {\bf 10}, 317, (1922).
\newline
[8]-- Fokker, A.D., Zeits. f. Physik, {\bf 58}, 386, (1929).
\newline
[9]-- Fokker, A.D., Physica, {\bf 9}, 33, (1929).
\newline
[10]-- Fokker, A.D., Physica, {\bf 12}, 145, (1932).
\newline
[11]-- Wheeler, J.A., and R.P. Feynman, Rev. Mod. Phys., {\bf 17}, No. 2,3, 157, (1945).
\newline
[12]-- Wheeler, J.A., and R.P. Feynman, Rev. Mod. Phys., {\bf 21}, No. 3, 425, (1949).
\newline
[13]-- Hoyle, F., and J.V. Narlikar, Proc. Roy. Soc. Lond., {\bf 277A}, 1, (1963).
\newline
[14]-- Hoyle, F., and J.V. Narlikar, Ann. Phys., {\bf 54}, 207, (1969).
\newline
[15]-- Hoyle, F., and J.V. Narlikar, Ann. Phys., {\bf 62}, 44, (1971).
\newline
[16]-- Hoyle, F., and J.V. Narlikar, Mon. Not. Roy. Astron. Soc., {\bf 155}, 323, (1972).
\newline
[17]-- Narlikar, J.V., Proc. Camb. Phil. Soc., {\bf 64}, 1071, (1968).
\newline
[18]-- Hoyle, F., and J.V. Narlikar, {\it Action at a distance in physics and cosmology\/},
W.H. Freeman and Company, San Francisco, (1974).
\newline
[19]-- Bohm, D., Phys. Rev., {\bf 85}, No. 2, 166, (1952). 
\newline
[20]-- Bohm, D., Phys. Rev., {\bf 85}, No. 2, 180, (1952). 
\newline
[21]-- Bohm, D., Phys. Rev., {\bf 89}, No. 2, 458, (1953).
\newline
[22]-- Bohm, D., B.J. Hiley, and P.N. Kaloyerou, Phys. Rep., {\bf 144}, No. 6, 321, (1987). 
\newline
[23]-- Bohm, D., B.J. Hiley, and P.N. Kaloyerou, Phys. Rep., {\bf 144}, No. 6, 349, (1987). 
\newline
[24]-- Bohm, D., and B.J. Hiley, {\it The undivided universe\/}, Routledge, London, (1993). 
\newline
[25]-- Holland, P.R., {\it The quantum theory of motion\/}, Cambridge University Press, (1993).
\newline
[26]-- Shojai, A., and M. Golshani, unpublished.
\newline
[27]-- Bell, J.S., Physics, {\bf 1}, 195, (1965).
\newline
[28]-- Clauser, J.F., and A. Shimony, Rep. Prog. Phys., {\bf 41}, 1881, (1978).
\newline
[29]-- de-Broglie, L., {\it Non-linear wave mechanics\/}, Elsevier Publishing Company, (1960).
\newline
[30]-- Squires, E.J., Phys. Lett. A., {\bf 178}, 22, (1993).
\newline
[31]-- Mackman, S., and E.J. Squires, Found. Phys., {\bf 25}, No. 2, 391, (1995).
\end{bf}
\end{document}